\documentclass[]{raa}            

\usepackage{graphicx,times}             
\usepackage{natbib}
\usepackage{amssymb,amsmath}
\usepackage{float}
\usepackage{booktabs}

\bibpunct{(}{)}{;}{a}{}{,}

\usepackage[pagebackref=true]{hyperref}

\begin{document}

  \title{
  Detecting HI Galaxies with Deep Neural Networks in the Presence of Radio Frequency Interference
  }

   \volnopage{Vol.0 (20xx) No.0, 000--000}      
   \setcounter{page}{1}          

   \author{Ruxi Liang 
      \inst{1,2}
   \and Furen Deng
      \inst{1,2}
   \and Zepei Yang
      \inst{3}
    \and Chunming Li
      \inst{4}
   \and Feiyu Zhao
      \inst{2,5}
    \and Botao Yang
      \inst{4}
    \and Shuanghao Shu
      \inst{1,2}
    \and Wenxiu Yang
      \inst{1,2}
    \and Shifan Zuo
      \inst{1,2}
    \and Yichao Li
      \inst{6}  
    \and Yougang Wang  \thanks{E-mail:wangyg@bao.ac.cn}
      \inst{1,2,6} 
    \and Xuelei Chen  \thanks{E-mail:xuelei@bao.ac.cn}
      \inst{1,2,6}
   }
   \institute{National Astronomical Observatories, Chinese Academy of Sciences, Beijing, 100101, China\\
        \and
             University of Chinese Academy of Sciences, Beijing, 101408, China\\
        \and
             College of Science, Northeastern University, Boston, MA, 02115, USA\\
        \and
             Biomedical Instrument Institute, School of Biomedical Engineering, Shanghai Jiao Tong University, Shanghai, 200030, China\\
        \and
             Shanghai Astronomical Observatory, Chinese Academy of Sciences, 80 Nandan Road, Shanghai, 200030, China\\
         \and 
             Key Laboratory of Cosmology and Astrophysics (Liaoning), College of Sciences, Northeastern University, Shenyang 110819, China\\  
\vs\no
   {\small Received 20xx month day; accepted 20xx month day}}

\abstract{In neutral hydrogen (HI) galaxy survey, a significant challenge is to identify and extract the HI galaxy signal from 
observational data contaminated by radio frequency interference (RFI). For a drift-scan survey, or more generally a survey of a spatially continuous region, in the time-ordered spectral data, the HI galaxies and RFI all appear as regions which extend an area in the time-frequency waterfall plot, so the extraction of the 
HI galaxies and RFI from such data can be regarded as an image segmentation problem, and machine learning methods can be applied to solve such problems. In this study, we develop a method to effectively detect and extract signals of HI galaxies based on a Mask R-CNN network combined with the PointRend method. By simulating FAST-observed galaxy signals and potential RFI impacts, we created a realistic data set for the training and testing of our neural network. 
We compared five different architectures and selected the best-performing one. This architecture successfully performs instance segmentation of HI galaxy signals in the RFI-contaminated time-ordered data (TOD), achieving a precision of 98.64\% and a recall of 93.59\%.
}

    \date{2023.4.24}
    \titlerunning{Detecting HI Galaxies with Deep Neural Networks}  

   \maketitle

%
%
\section{Introduction}           
\label{sect:intro}

    In recent years, a number of advanced radio telescopes and arrays have been constructed, including the Five-hundred-meter Aperture Spherical radio Telescope (FAST; \citealt{nanrd2011fast}), the Australian Square Kilometre Array Pathfinder (ASKAP; \citealt{Johnston2008ASKAP}), and MeerKat \citep{booth2012meerkat}, among others. 
    In the coming decade, the next generation of radio telescope arrays, such as the Square Kilometre Array (SKA; \citealt{Dewdney2009ska}), are anticipated to be completed. 
    The study of neutral hydrogen is one of the primary scientific goals of these telescopes, and HI galaxy surveys are key observations of them \citep{tolley2022lightweight}. 
    From the HI galaxy survey data, we can examine the HI content and mass function of the galaxies, gas accretion, the correlation between HI and star formation, and the influence of the environment on HI \citep{giovanelli2016extragalactic}. 
    These sensitive and precise instruments demand more sophisticated observational techniques and signal-processing methods.
    
    The HI Parkes All-Sky Survey (HIPASS; \citealt{Meyer2004HIPASS}) and the Arecibo Legacy Fast ALFA Survey (ALFALFA; \citealt{Giovanelli2005ALFALFA, Jones2018ALFALFA}) are the most extensive HI surveys completed so far. The `multifind' peak-finding algorithm \citep{kilborn2001multifind} was employed to identify and filter data signal peaks in the HIPASS data processing. This method searches local maxima in data cubes and identifies potential signals by setting a threshold. 
    The ALFALFA survey used a matched filtering algorithm \citep{Saintonge2007alfadata}, 
    which is sensitive to wide and weak signals.  Although these algorithms have served these surveys successfully, they still exhibit some shortcomings. 
    The {\tt multifind} result is sensitive to the threshold, and has difficulty with overlapping signals, or signals with unusual shapes and features. 
    The matched filtering algorithm also relies on assumptions about signal shapes, necessitating adjustments to algorithm parameters based on extensive experimentation and experience, it is prone to false alarms and missed detections when encountering multiple local maxima. The identification of radio frequency interference (RFI) is also far from perfect for these algorithms. More advanced and robust signal extraction methods are needed for future HI surveys.
    
    The RFI is always a challenge for radio astronomical observations face. RFI sources can be artificial or natural, with the former including digital television, mobile and satellite communications, and so on \citep{Fridman2001rfi}. 
    Efficient RFI mitigation algorithms which can identify the RFI are essential for radio astronomical observations. 
    Many automatic RFI flagging algorithms have been developed, typically by looking for unusually large deviations in the sample. For example, the widely used Sum-Threshold method \citep{offringa2010postrfi} searches RFI of different possible time and frequency spread by scanning the data with a sliding window, and comparing the sum of the power of consecutive samples with a blocksize-dependent threshold. 
    
    RFI mitigation and celestial signal extraction are two sides of the same process. In the past and present HI observations, the usual practice is to first remove the various interferences, including standing waves and RFI, through a pipeline, and then extract the desired HI signal from the processed data. However, the identification of RFI is not absolute, and the extraction process still faces the influence of some interference. Moreover, RFI often superimposes HI signals, causing contamination and rendering the data unusable. Therefore, a major challenge in the data processing of HI galaxy surveys is to identify the extragalactic HI signals amidst a vast amount of data.
    
    In recent years, there is much advancement in machine learning (ML), and it has been applied to various research directions in astronomy \citep{Ball2010ML}. In particular, these techniques have been applied to radio astronomical data processing tasks, such as RFI identification and mitigation, celestial source detection and classification, and analysis of observational data, among others \citep{baron2019machine}. Numerous deep learning-based models have been applied to identify and mitigate RFI, especially the Convolutional Neural Networks (CNN) \citep{sun2022robust, pinchuk2022machine}, U-Net \citep{akeret2017radio, yang2020deep}, and so on. Other ML-based image processing models have also been applied to astronomy, such as the new source finder developed by \cite{riggi2023astronomical} based on the Mask R-CNN framework for detecting and classifying sources in radio continuum images.
    
    Mask R-CNN is a CNN-based object detection and instance segmentation framework, which has achieved remarkable results in the field of computer vision \citep{he2017mask}. PointRend is a technique for improving image segmentation results by adding a rendering approach on top of the existing network, presenting fine-grained object boundaries through adaptive point sampling and label estimation \citep{kirillov2020pointrend}. This technique enhances segmentation quality, producing more refined edges.
    
    Inspired by these works, we apply the Mask R-CNN model and PointRend method to HI signal extraction in radio telescope data processing, hoping to more accurately detect and segment target objects in astronomical images. We develop an HI galaxy-searching method based on the Mask R-CNN model and the PointRend method. The model can directly search for and identify HI galaxies in time order data contaminated by RFI, and can extract signals by segmenting the data. Using FAST as an example, we simulated the observed HI galaxies and potential RFI impacts, and then trained, refined, and selected different architectures of PointRend Mask R-CNN models, ultimately achieving a good performance in identifying galaxy signals. 
    
    The structure of this paper is as follows. Sec. \ref{sect:Method} of the paper introduces the machine learning methods we used, including the principles of Mask R-CNN and PointRend, the network structure we employed, and the model evaluation method. 
    In Sec. \ref{sec:data}, the data preparation process is expounded. 
    Sec. \ref{sec:training} presents the training and testing of the networks, while Sec. \ref{sec:results} presents the final results of our experiment.
    Sec. \ref{sec:discuss} provides further analysis and discussion of the results. Finally, Sec. \ref{sec:conc} summarizes the entire paper.

\section{Method}
\label{sect:Method}

    \subsection{Machine Learning Method: Mask R-CNN and PointRend}
    \label{subsec:mlmethod}

        In this study, we develop the Mask R-CNN network by integrating it with the PointRend method 
        to accomplish the instance segmentation task of identifying HI galaxies in astronomical observation data.
        
        Mask R-CNN is an improved version of Faster R-CNN, which is a classic two-stage object detection network. 
        The Faster R-CNN represents detected objects by generating bonding boxes and the corresponding class information \citep{ren2016faster}.
        Mask R-CNN adds to the Faster R-CNN network a mask branch, which could generate the binary mask for each detected object. The additional mask branch significantly improves the network performance in instance segmentation tasks.
        
        The PointRend method is an innovative strategy that can be integrated with various neural networks. 
        By adding a PointRend head to the network, it improves the accuracy and resolution of image segmentation \citep{kirillov2020pointrend}. 
        After obtaining a preliminary coarse mask through other networks, PointRend generates some sampling points concentrated in the areas where the segmentation results are uncertain,  
        then adopts a sub-network called PointRend head to predict the classification of these points based on the input feature map. The predicted information is then combined with the coarse mask to generate a more precise mask.
    
        \begin{figure}[ht]
            \centering
            \includegraphics[width=\linewidth]{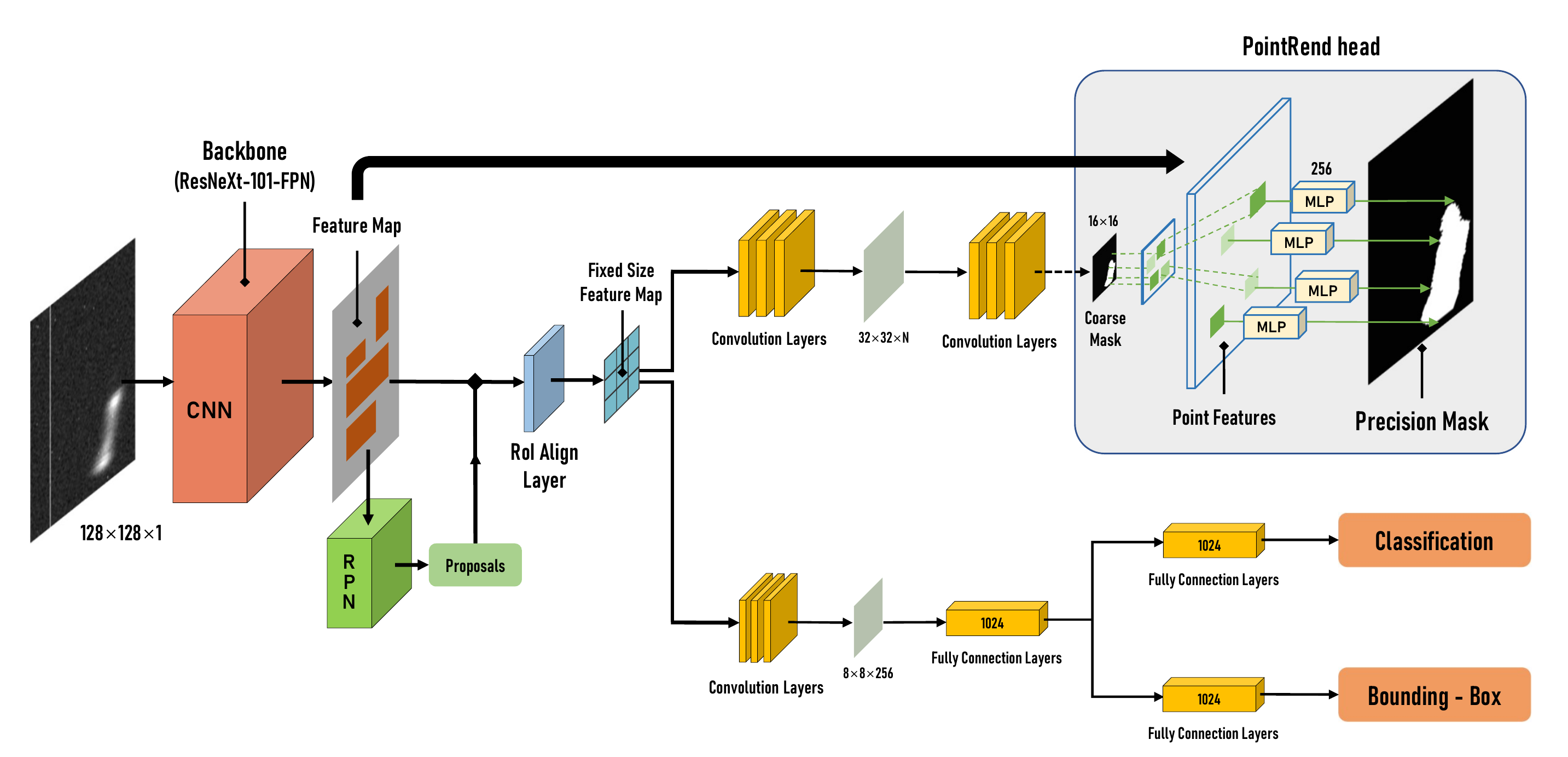}
            \caption{Schematic diagram of the PointRend Mask R-CNN model architecture.}
            \label{fig:PointRendstructure}
        \end{figure}
    
        Figure \ref{fig:PointRendstructure} shows the network structure of the model we used. For single-channel two-dimensional data, the model first obtains a feature map through the backbone network, which serves as the input for the Region Proposal Network (RPN; \citealt{ren2016faster}) and the final PointRend Head. 
        The RPN generates a series of proposals from the input feature map, each with a specific region where the target object might be located. 
        Each proposal is combined with the feature map to generate a series of RoIs (Regions of Interest), which are then passed through the RoI Align Layer \citep{he2017mask} to obtain fixed-size feature maps of uniform size. 
        There are two branches next. The branch at the bottom of the structure diagram is the RoI head branch, which has the same structure as the corresponding part of Faster R-CNN. 
        It first transforms the fixed-size feature map into a series of smaller maps by going through convolutional layers, and then obtains the class and bounding box information through several fully connected layers. This branch essentially completes the object detection of the input data.
        The upper branch is the Mask branch. After passing the fixed-size feature map through a series of convolutional layers, we obtain a coarse mask.
    
        Next is the PointRend part of the network. PointRend uses a sampling strategy based on uncertainty to generate some sampling points for refinement according to the coarse mask information. The model employs an additional sub-network called the PointRend Head, which receives the selected refinement points and the high-resolution feature map generated at the beginning of the entire network as input, and predicts the classification of each sampling point through a series of MLPs (Multi-Layer Perception). 
        Finally, the predicted class information is combined with the coarse mask to obtain a more accurate final precision mask. In each iteration, the sub-network calculates the uncertainty of the class prediction for each point, and selects a certain portion of points for updating based on the uncertainty values. This makes the selected points mainly located in the detail areas of the segmentation result (i.e., the edge areas and texture-complex areas), which are the areas that need improvement the most in the segmentation results.
    
        By undergoing these processes for each RoI, we can complete the instance segmentation of all targets in the input data.
        All parts of the model participate in training. By minimizing the loss function, the model updates the parameters of each network through backpropagation. We define the multi-task loss on each sampled RoI as
        \begin{align}
            \label{lossfunc}
            L_{\rm total} = L_{\rm cls} + L_{\rm box} + L_{\rm mask} + L_{\rm PointRend}.
        \end{align}
         $L_{\rm cls}$ is the classification loss defined as
        \begin{align}
            L_{\rm cls} = -\log(p_u),
        \end{align}
        where $p_u$ is the probability of an RoI belonging to the true class label $u \ (u \in \{1,2,\cdots,C \})$, calculated by the softmax function:
        \begin{align*}
            p_{u} = \frac{\exp(z_{u})}{\sum_{k=1}^{C}\exp(z_{k})},
        \end{align*}
        where $z_{i}$ represents the score of the RoI belonging to the $i$-th class for $i \in \{1,2,\cdots,C \}$.
        $L_{\rm box}$ is the bounding-box regression loss, which describes the bounding-box branch's ability to localize objects during bounding-box regression, defined as:
        \begin{align}
            L_{\rm box} = \sum_{i \in \{x,y,w,h\} }{{\rm Smooth}_{L_1}(t_i^u - v_i)},
        \end{align}
        where $v = (v_x, v_y, v_w, v_h)$ is the true bounding-box regression targets for class $u$ and $t^u = (t^u_x, t^u_y, t^u_w, t^u_h)$ is the predicted bounding box coordinates. ${\rm Smooth}_{L_1}$ represents the Smooth $L_1$ loss function:
        \begin{align}
            {\rm Smooth}_{L_1}=\begin{cases}
                        0.5x^2      &		{\rm if}\ \left| x \right|<1\\
                        \left| x \right|-0.5&		{\rm otherwise}\\\end{cases}.
        \end{align}
        $L_{\rm mask}$ is described as the average binary cross-entropy per pixel and describes the mask head's ability to classify each pixel, defined as
        \begin{align}
            L_{\rm mask} = -\frac{1}{m^2}\sum_{i=1}^{m^2}[p_i\log(\hat{p_i}) + (1-p_i)\log(1-\hat{p_i})],
        \end{align}
        where $m^2$ is the total number of pixels in the mask ($16\times16$ in our network), $p_i$ is the true value of the $i$-th pixel, and $\hat{p_i}$ is the predicted value of the $i$-th pixel.
    
        The PointRend loss, $L_{\rm PointRend}$, calculates binary cross-entropy only on the sampled points that need to be refined:
        \begin{align}
            L_{\rm PointRend} = -\frac{1}{N_{\rm point}}\sum_{i=1}^{N_{\rm point}}[p_i\log(\hat{p_i}) + (1-p_i)\log(1-\hat{p_i})],
        \end{align}
        where $N_{\rm point}$ represents the number of sampled points that need refinement (set to 10 in our task), $p_i$ is the true value of the $i$-th point, and $\hat{p_i}$ is the predicted value of the $i$-th point.
        In addition, the RPN network used in the model has its own loss function and is trained independently during the training process.

    \subsection{Model Evaluation}
    \label{subsec:modelevalue}
    
        In our PointRend Mask R-CNN network, we selected five distinct backbones for obtaining feature maps and conducted a comparative analysis to evaluate the impact of different backbones on the network performance. Our choices are all based on Residual Networks (ResNet; \citealt{he2015deepresidual}), which is a deep convolutional neural network. 
        The ResNet introduces residual connections to solve the problem of gradient vanishing and explosion issues during the training of deep neural networks.
    
        In our selection, ResNet-50-FPN and ResNet-101-FPN are Feature Pyramid Networks (FPN; \citealt{lin2017featureFPN}) based on 50-layer and 101-layer residual networks, respectively. 
        These FPNs add a top-down pathway and lateral connections to the original ResNet, enabling the network to better capture features at different scales, and could improve object detection and instance segmentation performance by leveraging multi-scale features. 
        ResNet-50-C5-Dilated and ResNet-101-C5-Dilated are dilated convolutional networks based on 50-layer and 101-layer residual networks, respectively, using dilated convolution in the last convolutional layer (C5) \citep{yu2016multiscaleDilated}. 
        This approach increases the receptive field size, thereby improving the detection and segmentation performance for large-scale objects. 
        Our primary focus is on the ResNeXt-101-FPN backbone. ResNeXt-101 is an improved 101-layer ResNet network that employs grouped convolution on top of ResNet, dividing the input channels into multiple groups and performing convolution operations within each group. This enhances the network's expressiveness and parameter efficiency, allowing for improved performance with relatively low computational complexity \citep{xie2017aggregatedResNeXt}. 
        After combined with FPN, ResNeXt-101-FPN should perform slightly better than ResNet-101-FPN theoretically.
        
        Generally, deeper network structures can usually learn more diverse feature representations, thereby improving the accuracy of instance segmentation. 
        FPN can effectively capture multi-scale feature information by integrating features from different levels, resulting in better performance when dealing with objects of varying sizes. Dilated convolution, by expanding the receptive field of the convolution kernel, can better capture information from large-scale objects. For our project, an FPN with a deeper structure is theoretically more suitable.
    
        We employed the Precision, Recall rate, and F1 Score, which are commonly used performance metrics for evaluating image segmentation models (\citealt{Forsyth2002ComputerVA}), to evaluate our method. The Precision in this case is the proportion of true galaxies among the samples classified as galaxies by the model, reflecting the accuracy of the model in recognizing galaxies. 
        \begin{align}\label{eq:p}
            {\rm Precision} = \frac{\rm TP}{\rm TP+FP},
        \end{align}
        where TP represents a correct segmentation (detection) that the instance is classified as a member of the class while
        FP represents an incorrect segmentation of such classification. Precision belongs to $[0,1]$, and a higher value indicates that the model is less likely to misidentify. 
        
        The Recall rate in the present case is the fraction of identified HI galaxies among all HI galaxies, representing the model's capability of detection. Recall belongs to $[0,1]$, and a higher value indicates a stronger recognition ability. It is defined as
        \begin{align}
            {\rm Recall} = \frac{\rm TP}{\rm TP+FN},
        \end{align}
        where FN represents an incorrect segmentation that the instance is not classified as a member of the class.
        
        The F1 Score is the harmonic mean of the Precision and the Recall, also belonging to $[0,1]$, providing a comprehensive evaluation of both Precision and Recall performance. It can serve as the standard for assessing the model, and a higher value indicates the model has a better performance. It is defined as
        \begin{align}
            {\rm F1} = \frac{2\times{\rm Precision}\times {\rm Recall}}{{\rm Precision}+{\rm Recall}}.
        \end{align}

\section{Mock Data}
\label{sec:data}

    The PointRend Mask R-CNN is a supervised neural network model that requires data for training and testing. For our mission, the construction of datasets can be diverse. Referring to the FAST telescope, we simulated the HI galaxy signals it could observe and the possible RFI effects it might encounter.

    \subsection{HI Galaxies Data Simulation}
    
        We generate the mock HI galaxies from the IllustrisTNG magnetohydrodynamical \citep{Weinberger2017sim, Pillepich2018TNG} simulation. It includes physical processes such as gas cooling, star formation, stellar evolution, metal enrichment, black hole growth, stellar winds, supernovae, and active galactic nuclei (AGNs), and is consistent with current observations. In this work, we use the TNG100-1 data set. The box size for TNG100 is $75 \  {\rm Mpc}/h$ and mass resolution is $9.4\times 10^5\ M_\odot/h$ for baryon particle and $5.1\times 10^6\ M_\odot/h$ for dark matter particle. The box size ensures that there are enough galaxies as training sets and test sets, and the mass resolution ensures that the structure of galaxies can be resolved.  The IllustrisTNG adopted the Planck 2015 cosmological parameters \citep{ade2016planck}, i.e., $\Omega_\Lambda=0.6911$, $\Omega_m=0.3089$, $\Omega_b=0.0486$, $\sigma_8=0.8159$, $n_s=0.9667$, and $h=0.6774$, and we adopt this model throughout the paper. 
        
        The total mass of atomic and molecular hydrogen for each gas particle can be obtained directly from the TNG100 catalog. However, these two parts are not separated in the simulation. \cite{diemer2018modeling} has separated the molecular and atomic hydrogen contents for galaxies in TNG100. However, we also need to take into account the velocity of each gas particle to get the spectral profile for each galaxy, which can not be obtained from the existing catalog. We calculate the HI mass for each gas particle based on the method of \cite{gnedin2011environmental}. We refer readers to \cite{Furen2022forecast} for details of the calculation. 
        Following \cite{diemer2018modeling} we consider only galaxies with stellar mass or gas mass greater than $2\times 10^8\ M_\odot$, which are well represented by particles in this simulation.  
                
\begin{figure}[htbp]
            \centering
            \includegraphics[width=0.7\linewidth]{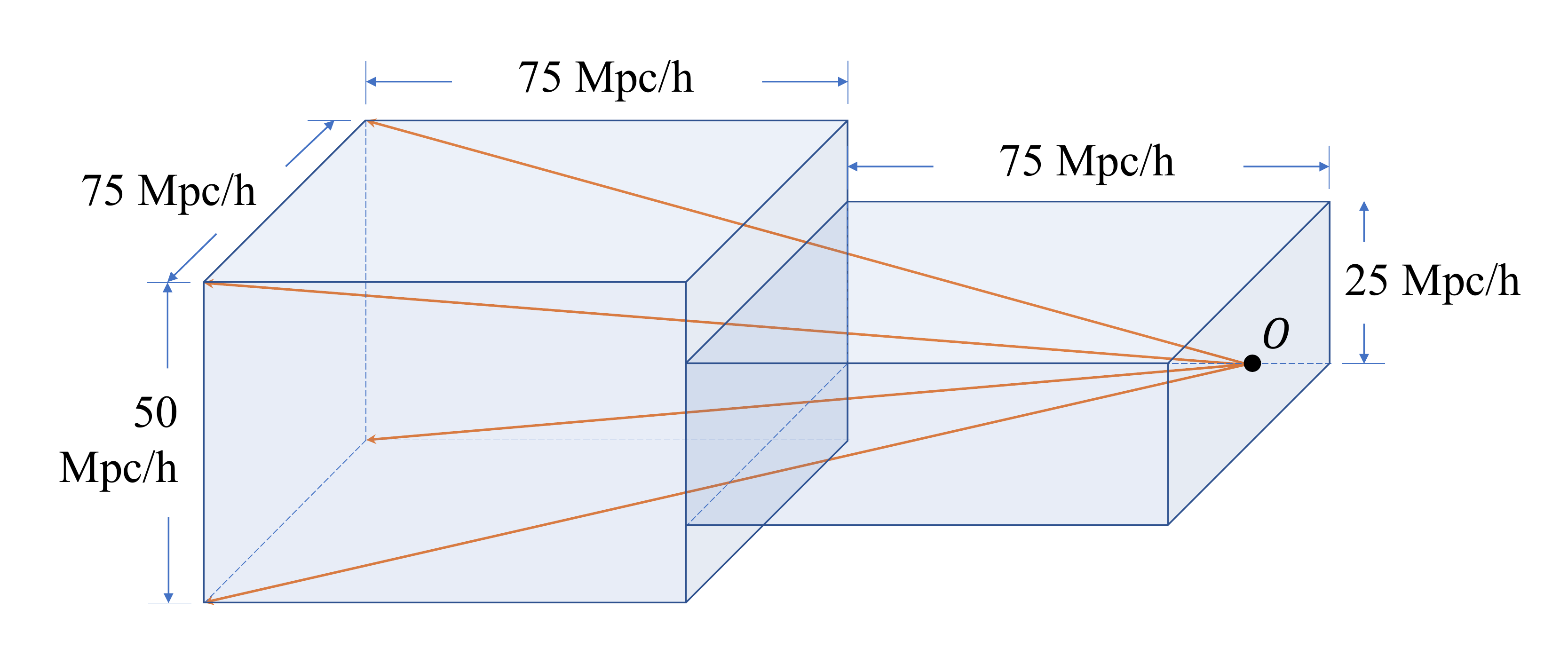}
            \caption{The configuration of the light cone by stacking two boxes. The size of two boxes is shown in the figure, $O$ is the observer and the orange lines join the observer with the four corners of the field of view. The redshift range of the light cone extends to $z\approx 0.05$, and its angular area is approximately $28\times 19\ {\rm deg^2}$.}
            \label{fig:lightcone}
        \end{figure}

        We assume that the properties of HI galaxies do not evolve significantly over the small redshift range considered in this work, and only use the simulation snapshot at $z\approx 0$. The box is split into two boxes with size $75 \times 75 \times 50\ ({\rm Mpc}/h)^3$ and $75 \times 75 \times 25\ ({\rm Mpc}/h)^3$. Then we stack the two boxes to form a light cone volume as shown in Figure \ref{fig:lightcone}, where $O$ is the observer. We have a rectangular field of view and the orange lines join the observer with the four corners of the field. 
        We choose this configuration to ensure sufficient redshift coverage and field of view while avoiding the repeating of galaxy samples.
        The redshift range of the light cone extends to $z\approx 0.05$, and its angular area is approximately $28\times 19\ {\rm deg^2}$.
        We then deposit the gas particles into angular and frequency grids, where the angular grid has a size of $\Delta\theta=0.5\ {\rm arcmin}$, well below the beam resolution of FAST, and the frequency grid has a size of $\Delta \nu=0.02\ {\rm MHz}$, to suit the purpose of the galaxy detection. 
        The frequency of each gas particle is determined as $\nu=\nu_{21}/(1+z)/(1 + \beta)$, where $\nu_{21}\approx 1420.4\ {\rm MHz}$ is the rest-frame frequency of 21 cm radiation, $z$ is the cosmological redshift, and $\beta$ is the line-of-sight component of peculiar velocity in units of the speed of light.

\begin{figure}[htbp]
            \centering
            \includegraphics[width=0.5\linewidth]{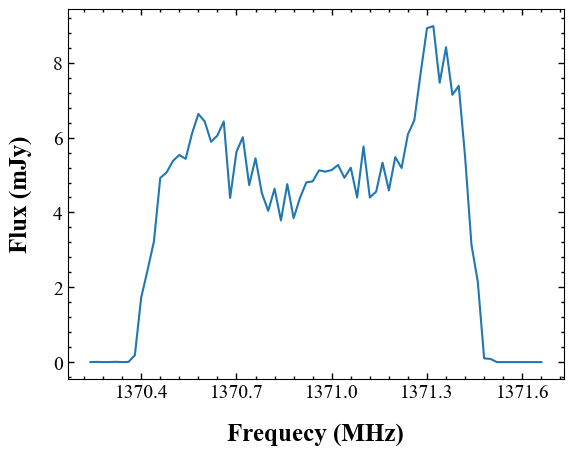}
            \caption{The spectrum of one of our simulated galaxies. 
            The peak flux is about $8\ {\rm mJy}$ and the line width is about $1\ {\rm MHz}$ with a characteristic `double horn' profile.}
            \label{fig:spectrum1}
        \end{figure}

        We calculate the brightness temperature for cell $i$ in frequency $\nu$ by
        \begin{equation}
            T_b^i(\nu) = \frac{3c^2}{32\pi\nu_{21}^3} A_{10} \frac{h\nu^2}{k_B m_p} \frac{\Delta M_{\rm HI}^i}{D_A(z)^2\Delta\theta^2\Delta \nu},
        \end{equation}
        where $c$ is the speed of light, 
        $A_{10}\approx 2.85\times 10^{-15}\ {\rm Hz}$ is the spontaneous emission coefficient of the 21 cm transition, 
        $m_p$ is the mass of the proton, 
        $k_B$ is the Boltzmann constant,
        $D_A(z)$ is the angular diameter distance,
        and $\Delta M_{\rm HI}^i$ is the HI mass in cell $i$.
        We ignored the velocity dispersion inside the gas particle in our calculation which may smooth the spectrum but cannot be obtained from the simulation. The spectrum of one simulated galaxy is shown in Figure \ref{fig:spectrum1}. Its peak flux is about $8\ {\rm mJy}$ and the line width is about 1 MHz with a characteristic `double horn' profile. It is consistent with our knowledge about the HI profile in low redshift galaxies (\citealt{Saintonge2007alfadata}).
        
        We model the beam of FAST as a Gaussian function with $\sigma = 0.518 \lambda/(300 \ {\rm m})$, though the real beam may have a more complicated dependence on the frequency. The 19 beams are rotated $23.4^\circ$ w.r.t the configuration given in \cite{Jiangpeng2020FAST}, to achieve a more uniform coverage in drift scan. We place the angular center of the grids at the zenith, and assume the sky is surveyed with the rotation of the Earth. 
        The scan produces strips along the right ascension (RA) direction. According to \cite{Jiangpeng2020FAST}, the 19 beams cover over 25 arcmin in the direction of declinations (DEC), so by repeatedly scanning different declinations with a separation of $10\ {\rm arcmin}$, the whole available angular area (which is approximately $28\times 19\ {\rm deg^2}$) is surveyed.
        We set the time resolution as $1.00663296\ {\rm s}$ and frequency resolution as $0.02\ {\rm MHz}$.

        \begin{figure}[htbp]
            \centering
            \includegraphics[width=\linewidth]{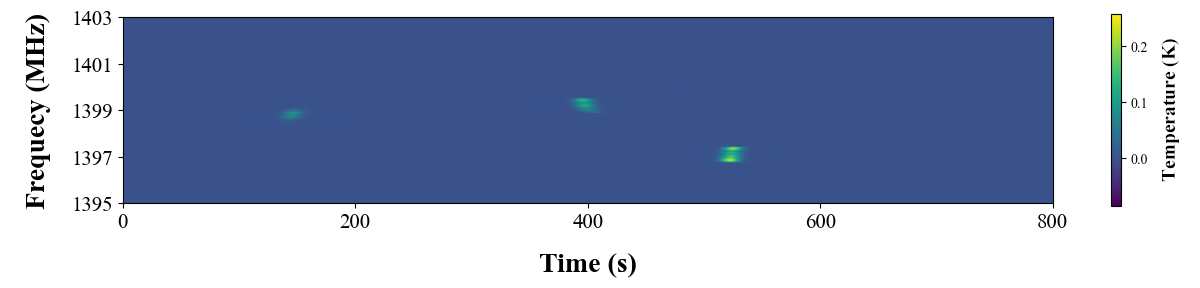}
            \caption{A piece of simulated TOD data of HI galaxies, with the horizontal axis representing time, the vertical axis representing frequency, and the color representing the antenna temperature value. As one can see, there are three galaxies present in the figure.}
            \label{fig:xingxi1}
        \end{figure}

        Based on the sensitivity of FAST, we dropped the galaxies with fluxes below $5\ {\rm mJy}$. Additionally, when training with data from a single beam, we also removed those galaxies that could be observed by other beams but were not visible to the specific beam we use.
        Ultimately, we obtained 4495 HI galaxies. These galaxies include both bright and faint ones, with varying shapes, and produce different brightness levels in the time order data, essentially covering various scenarios encountered in real observations. Figure \ref{fig:xingxi1} illustrates a piece of simulated HI galaxy signals, containing only galaxy signals without RFI, allowing us to label each galaxy easily and conveniently.

    \subsection{RFI Simulation}
    
        There are a number of software packages that can be employed for simulating RFI. The {\tt HIDE} (HI Data Emulator) is a software package for simulating HI observation data, and it could also generate mock RFI \citep{akeret2017hide, yang2020deep}. The {\tt Hera\_sim} \citep{Kerrigan2019herasim} is a Python software package developed for simulating the Hydrogen Epoch of Reionization Array (HERA) data, which can also generate RFI data \citep{sun2022robust}.  We integrated and adapted these two software packages to simulate RFI. We considered several types of RFI, including narrowband RFI, broadband RFI, and `clump' RFI.

\begin{figure}[ht]
            \centering
            \includegraphics[width=\linewidth]{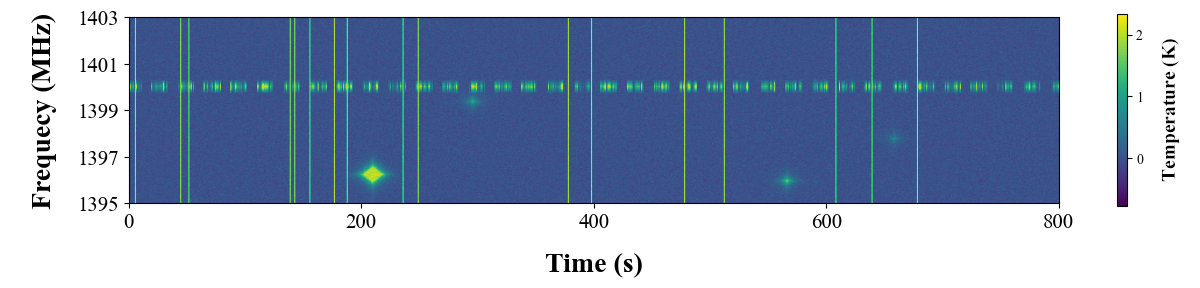}
            \caption{A piece of simulated TOD data of RFI, with the horizontal axis representing time, the vertical axis representing frequency, and the color representing the antenna temperature value. Narrowband RFI manifests as discontinuous horizontal lines with width, broadband RFI appears as thin vertical lines and stain-like RFI presents as large and small radiating spots.}
            \label{fig:RFI}
        \end{figure}

        Broadband RFI is instantaneous and intense, typically originating from sources such as lightning and transmission cables, generally covering many frequency bands and manifesting as `bright lines' spanning numerous frequencies in time-ordered data. Narrowband RFI is usually caused by digital television, satellite communications, and mobile communication. A typical narrowband RFI appears as a long-lasting and narrow frequency spread signal, presenting as intermittent stripes in time-ordered data. Another type of RFI, with a frequency spread and appearance time similar to galaxies, may stem from harmonics of satellite communications and certain short-term electromagnetic wave emissions. It exhibits a stain-like clump shape in time-ordered data and is more prone to confusion with galaxies. Ultimately, we successfully simulated these types of RFI. We then generated system noise following the method in \cite{Jiangpeng2020FAST} and added it to the data, Figure \ref{fig:RFI} shows a segment of the mock RFI and noise data, displaying different types of RFI.

\section{Model Training and Testing}
\label{sec:training}
    
    With the mock data generated above, we trained our network model. We divided the dataset into a training set, a validation set, and a test set with a 3:1:1 ratio, then we trained and tested the PointRend Mask R-CNN model with five different backbones.

    We set the batch size to 16, and the maximum number of iterations to 50,000. Each training sample generates classes, bounding-boxes, and mask predictions after training. The network also parses the classes and bounding-boxes information from the true mask, which serves as the ground truth. By calculating and minimizing the loss function value according to the method presented in Sec. \ref{subsec:mlmethod}, the model parameters are updated through backpropagation, thereby training the model. We employed the SGDM (Stochastic Gradient Descent with Momentum, \citealt{Ning1999momentum}) method to update the parameters, which could help accelerate convergence, and set the momentum as 0.9 \citep{Sutskever2013OnTI}.
    

    \begin{figure}[ht]
        \centering
        \begin{subfigure}{0.7\textwidth}
            \centering
            \includegraphics[width=\linewidth]{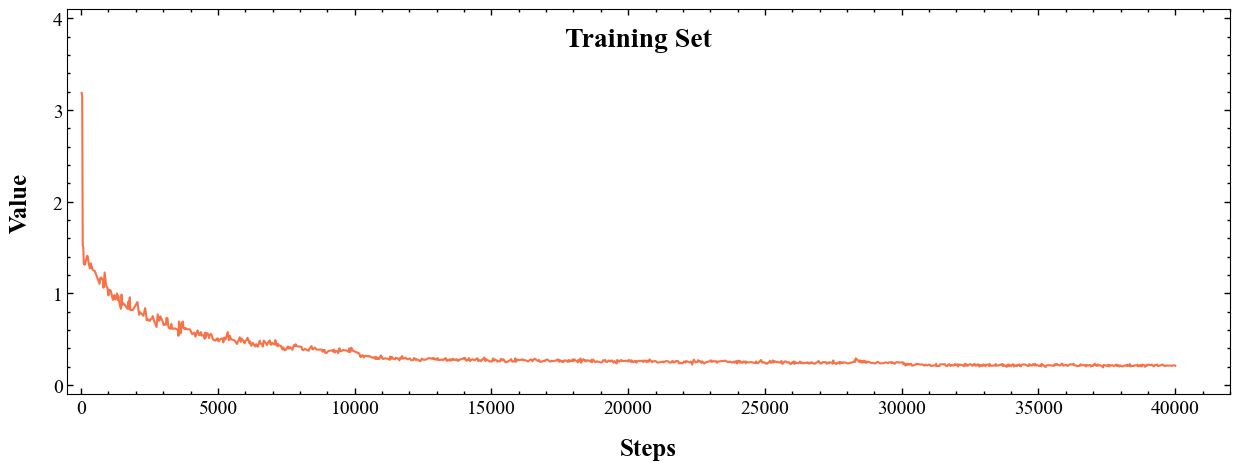}
        \end{subfigure}
        \hfill
        
        \vspace{1ex}
        
        \begin{subfigure}{0.7\textwidth}
            \centering
            \includegraphics[width=\linewidth]{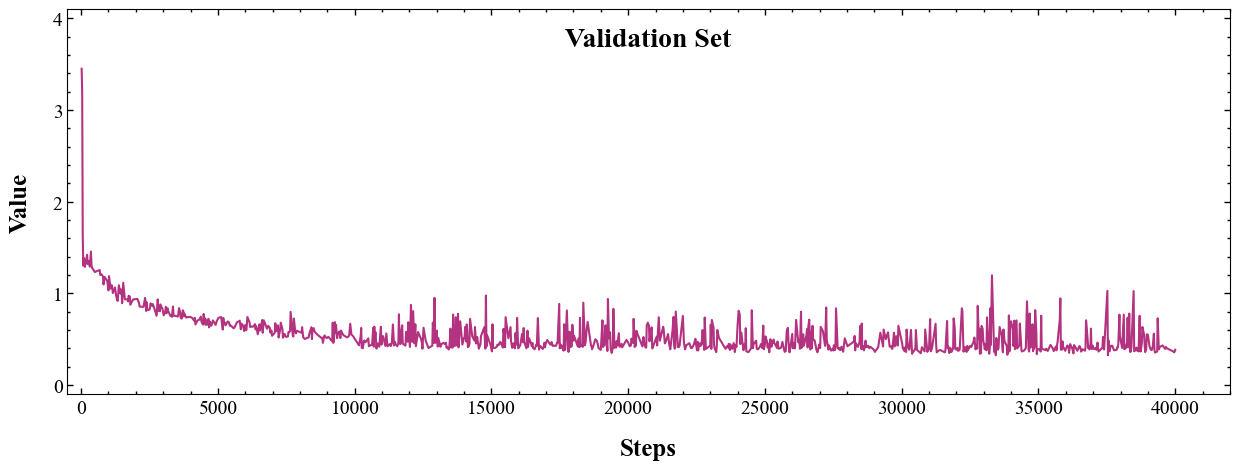}
        \end{subfigure}
        \caption{Loss function values of the model with ResNeXt-101-FPN backbone during the training process. The upper panel illustrates the variation of loss function values on the training set, while the lower panel shows the variation of loss function values on the validation set. The horizontal axis represents the training iteration steps, and the vertical axis indicates the function values. As can be seen, the loss function eventually converges on the training set. On the validation set, the loss function also exhibits a decreasing and converging trend. However, slight oscillations occur after 12,000 steps and intensify after 32,000 steps. Ultimately, we chose the model trained up to 32,000 steps.}
        \label{fig:loss}
    \end{figure}
    
    The base learning rate was set as 0.0005. We utilized a learning rate warmup strategy \citep{goyal2018accurate}, in which the learning rate will increase gradually and linearly from a lower value to the preset base learning rate during the initial stage of training. This strategy helps the model converge more stably in the early training phase and reduces the risk of gradient explosion. We also employed a multi-step learning rate decay strategy \citep{Krizhevsky2012ImageNetCW}, decaying the learning rate at the 10,000-th and 30,000-th iterations by multiplying the preset decay factor (set as 0.6) with the current learning rate. This strategy assists in fine-tuning the model parameters during the later stages of training, providing more precise adjustments as the training progresses, thereby enhancing the model's performance. Both strategies have been widely used in deep learning. Weight decay was employed as a regularization technique to prevent overfitting and enhance the model's generalization capabilities, with the weight decay coefficient set as 0.0001. This is also a widely applied strategy \citep{Goodfellow-et-al-2016}.

    For the backbone of our main concern, ResNeXt-101-FPN, Figure \ref{fig:loss} demonstrates the variations of the loss function values on the training and validation sets during the training process. As one can see, the loss function exhibits a decreasing trend and eventually converges on the training set. On the validation set, the loss function also displays a general decreasing and converging trend. Although slight oscillations in the loss function values on the validation set appear after 12,000 steps and intensify after 32,000 steps, the function values do not have an increasing trend, indicating that our model does not exhibit significant overfitting. 
    Such oscillations are normal and can be attributed to the inherent randomness in the optimization process, and this intensification in the later stages of training may be related to the size of the validation set and the batch size settings.
    In our mission, if the model experiences underfitting, it may not effectively learn and recognize various galaxy signals within the data, resulting in a low recognition capability. In contrast, in the case of overfitting, the model could become overly focused on the features within the current training data, leading to a decline in generalization performance on new data. To minimize the occurrence of both underfitting and overfitting, we monitored the model's performance on the validation set and ensured that the oscillations are within an acceptable range. Ultimately, we chose the network trained to 32,000 steps as our model. Similar phenomena were observed in the training processes of the networks with other backbones, and we also selected the final model for each backbone at the training step where the loss function had relatively converged on the training set and before the intensification of oscillations on the validation set.
    
    After training, we tested the model using the test set and calculated the model evaluation metrics according to the method in Sec. \ref{subsec:modelevalue}.
    The model was trained on NVIDIA GeForce RTX 2080 Ti GPU.

\section{Results}
\label{sec:results}

    For the training results of the PointRend Mask R-CNN model with different backbones, we calculated their Precision, Recall, and F1 Score respectively, as shown in Table \ref{resulttable}.

    \begin{table}[H]
        \begin{center}
        \caption[]{Precision, Recall and F1 Score of Our PointRend Mask R-CNN Network with Different Backbones.}\label{resulttable}
        \begin{tabular}{@{}clccc@{}}
        \toprule
                   & Backbone      & Precision  & Recall & F1 Score \\ \midrule
        PointRend Mask R-CNN  & ResNet-50-FPN  & 96.15\% & 94.93\%   & 95.54\% \\
                   & ResNet-50-C5-Dilated  & 100\% & 65.38\%   & 79.07\%  \\
                   & ResNet-101-FPN & 92.68\% & 97.43\%   & 95.00\%  \\
                   & ResNet-101-C5-Dilated  & 98.14\% & 67.94\%   & 80.29\%  \\
                   & ResNeXt-101-FPN & 98.64\% & 93.59\%   & 96.05\%  \\ \bottomrule
        \end{tabular}
        \end{center}
    \end{table}

    The ResNet-50-C5-Dilated and ResNet-101-C5-Dilated backbones performed well in terms of Precision, but their Recall is quite poor, leading to a low F1 score. The F1 score for the ResNet with FPN is generally better than that of the C5-Dilated ResNet. Moreover, the performance of ResNeXt-101-FPN is slightly better than that of the ResNet-50-FPN and ResNet-101-FPN, which is consistent with our expectations. Ultimately, we selected ResNeXt-101-FPN as the backbone for our model.

\begin{figure}[htbp]
        \centering
        \begin{subfigure}{0.32\textwidth}
            \centering
            \includegraphics[width=\linewidth]{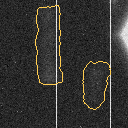}
        \end{subfigure}
        \hfill
        \vspace{0.6ex}
        \begin{subfigure}{0.32\textwidth}
            \centering
            \includegraphics[width=\linewidth]{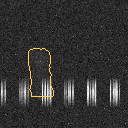}
        \end{subfigure}
        \hfill
        \vspace{0.6ex}
        \begin{subfigure}{0.32\textwidth}
            \centering
            \includegraphics[width=\linewidth]{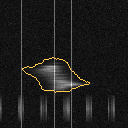}
        \end{subfigure}\\
        \vspace{0.6ex}
        \begin{subfigure}{0.32\textwidth}
            \centering
            \includegraphics[width=\linewidth]{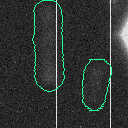}
            \caption{}
            \label{fig:good_1_GT}
        \end{subfigure}
        \hfill
        \vspace{0.6ex}
        \begin{subfigure}{0.32\textwidth}
            \centering
            \includegraphics[width=\linewidth]{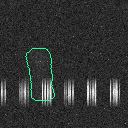}
            \caption{}
            \label{fig:good_2_GT}
        \end{subfigure}
        \hfill
        \vspace{0.6ex}
        \begin{subfigure}{0.32\textwidth}
            \centering
            \includegraphics[width=\linewidth]{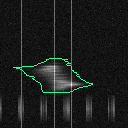}
            \caption{}
            \label{fig:good_3_GT}
        \end{subfigure}\\
        \vspace{0.6ex}
        \begin{subfigure}{0.32\textwidth}
            \centering
            \includegraphics[width=\linewidth]{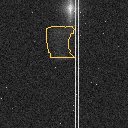}
        \end{subfigure}
        \hfill
        \vspace{0.6ex}
        \begin{subfigure}{0.32\textwidth}
            \centering
            \includegraphics[width=\linewidth]{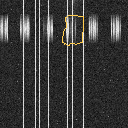}
        \end{subfigure}
        \hfill
        \vspace{0.6ex}
        \begin{subfigure}{0.32\textwidth}
            \centering
            \includegraphics[width=\linewidth]{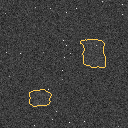}
        \end{subfigure}
        \vspace{0.6ex}
        \begin{subfigure}{0.32\textwidth}
            \centering
            \includegraphics[width=\linewidth]{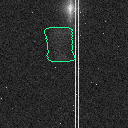}
            \caption{}
            \label{fig:good_4_GT}
        \end{subfigure}
        \hfill
        \begin{subfigure}{0.32\textwidth}
            \centering
            \includegraphics[width=\linewidth]{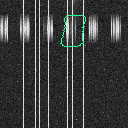}
            \caption{}
            \label{fig:good_5_GT}
        \end{subfigure}
        \hfill
        \begin{subfigure}{0.32\textwidth}
            \centering
            \includegraphics[width=\linewidth]{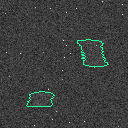}
            \caption{}
            \label{fig:good_6_GT}
        \end{subfigure}
        \caption{Examples of the HI galaxies correctly recognized by our model. These images are all plotted by part of our final simulated TOD data, with the horizontal axis representing time, the vertical axis representing frequency, and the brightness representing the value of the antenna temperature. In the images, a brighter (whiter) pixel represents a higher temperature at that point, with the brightest areas reaching about $3\ {\rm K}$. 
        Yellow lines delineate the galaxy contour determined by the model's output mask. The green lines represent the ground truth, which is the galaxy contour in our simulated TOD data. Other bright areas in the images correspond to various RFI and noise.}
        \label{fig:GoodResults}
    \end{figure}

    Figure \ref{fig:GoodResults} illustrates some examples of galaxies correctly recognized by the model, with the yellow lines delineating the regions determined by the model's output mask, and the green lines representing the ground truth.
    In Figure \ref{fig:good_1_GT}, there is a bright RFI spot on the right side, with one galaxy contaminated by broadband RFI, but the model still accurately identifies all the two galaxies, successfully detects multiple targets. 
    From Figure \ref{fig:good_2_GT}, one can see that our model can also effectively discern galaxy data contaminated by narrowband RFI. 

    When a bright galaxy is encountered, as illustrated in Figure \ref{fig:good_3_GT}, our model does not simply identify the `bright' regions, but also captures the faint areas at the edges of the galaxy data, indicating that the model has learned the characteristics of HI galaxies during training. From an image processing perspective, the high gradient at the edges of such bright galaxies could easily lead to overfitting during training, but our model does not exhibit this issue. Figure \ref{fig:good_4_GT} presents a typical situation where a galaxy and an easily confused RFI portion overlap. As can be seen, after sufficient training, the model can effectively handle such cases.

    The RFI contamination is more intense in Figure \ref{fig:good_5_GT}, affecting almost half of the galaxy's signal, yet the model still recognizes the galaxy, demonstrating its strong identification capability. Figure \ref{fig:good_6_GT} shows that the model can successfully identify faint galaxies with lower signal-to-noise ratios. 
    Each of Figure \ref{fig:good_1_GT} and Figure \ref{fig:good_6_GT} contains more than one HI galaxy target, highlighting the necessity of performing instance segmentation for galaxies.
    
    Considering the diverse morphology of galaxies and the variety of RFI patterns, our model demonstrates strong generalization capabilities, indicating that it can successfully accomplish the task of instance segmentation for finding HI galaxies in RFI-contaminated data.
    
\begin{figure}[htbp]
        \centering
        \begin{subfigure}{0.32\linewidth}
            \centering
            \includegraphics[width=\linewidth]{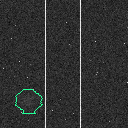}
            \caption{Groundtruth}
        \end{subfigure}
        \hfill
        \vspace{0.6ex}
        \begin{subfigure}{0.32\linewidth}
            \centering
            \includegraphics[width=\linewidth]{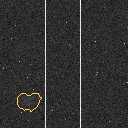}
            \caption{ResNeXt-101-FPN}
        \end{subfigure}
        \hfill
        \vspace{0.6ex}
        \begin{subfigure}{0.32\linewidth}
            \centering
            \includegraphics[width=\linewidth]{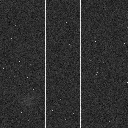}
            \caption{ResNet-50-C5-Dilated}
        \end{subfigure}
        \vspace{0.6ex}
        \begin{subfigure}{0.32\linewidth}
            \centering
            \includegraphics[width=\linewidth]{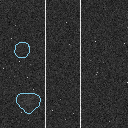}
            \caption{ResNet-50-FPN}
        \end{subfigure}
        \hfill
        \begin{subfigure}{0.32\linewidth}
            \centering
            \includegraphics[width=\linewidth]{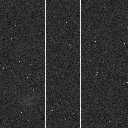}
            \caption{ResNet-101-C5-Dilated}
        \end{subfigure}
        \hfill
        \begin{subfigure}{0.32\linewidth}
            \centering
            \includegraphics[width=\linewidth]{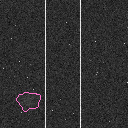}
            \caption{ResNet-101-FPN}
        \end{subfigure}
        \vspace{-0.4cm}
        \caption{An example from the results of our model with different backbones. The presentation way of these images is the same as in Fig. \ref{fig:GoodResults}, and each image represents the recognition results of the model with the corresponding backbone. Specifically, the green lines represent the galaxy contour determined by the ground truth, while the yellow, blue, and pink lines represent the galaxy contour determined by our model with ResNeXt-101-FPN, ResNet-50-FPN and ResNet-101-FPN backbones, respectively.}
        \label{fig:BadResults1}
    \end{figure}

\begin{figure}[htbp]
        \begin{subfigure}{0.32\linewidth}
            \centering
            \includegraphics[width=\linewidth]{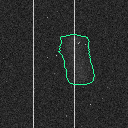}
            \caption{Groundtruth}
            \label{fig:bad_1_GT}
        \end{subfigure}
        \hfill
        \vspace{0.6ex}
        \begin{subfigure}{0.32\linewidth}
            \centering
            \includegraphics[width=\linewidth]{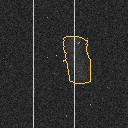}
            \caption{ResNeXt-101-FPN}
        \end{subfigure}
        \hfill
        \vspace{0.6ex}
        \begin{subfigure}{0.32\linewidth}
            \centering
            \includegraphics[width=\linewidth]{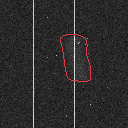}
            \caption{ResNet-50-C5-Dilated}
        \end{subfigure}
        \vspace{0.6ex}
        \begin{subfigure}{0.32\linewidth}
            \centering
            \includegraphics[width=\linewidth]{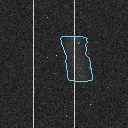}
            \caption{ResNet-50-FPN}
        \end{subfigure}
        \hfill
        \begin{subfigure}{0.32\linewidth}
            \centering
            \includegraphics[width=\linewidth]{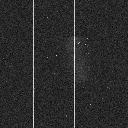}
            \caption{ResNet-101-C5-Dilated}
        \end{subfigure}
        \hfill
        \begin{subfigure}{0.32\linewidth}
            \centering
            \includegraphics[width=\linewidth]{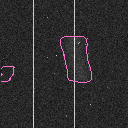}
            \caption{ResNet-101-FPN}
        \end{subfigure}
        \vspace{-0.4cm}
        \caption{Another example from the results of our model with different backbones. The presentation way of these images is the same as in Fig. \ref{fig:GoodResults}, and each image represents the recognition results of the model with the corresponding backbone. The meaning of lines in different colors is the same as in Fig. \ref{fig:BadResults1}, and the red lines represent the galaxy contour determined by our model with ResNet-50-C5-Dilated as the backbone.}
        \label{fig:BadResults2}
    \end{figure}

    We present the different recognition results of the same example using different backbones as well, with the outcomes of various models marked with distinct color contours. Figure \ref{fig:BadResults1} shows a rather faint galaxy, with the two ResNet models using C5-Dilated failing to detect the target. In contrast, ResNet-50-FPN produces a false detection, possibly due to the influence of some faint galaxies with a low signal-to-noise ratio in the training dataset, leading the model to misinterpret random noise fluctuations as galaxies. Figure \ref{fig:BadResults2} displays a galaxy contaminated by broadband RFI, with ResNet-101-C5-Dilated still missing the target, while ResNet-101-FPN produces a false detection. These examples illustrate that, in some cases, galaxy recognition can be quite challenging, and models may have limitations and areas for improvement.

    \vspace{0.3ex}
        
\section{Discussion}
\label{sec:discuss}

    \subsection{Construction of the Dataset}
    
        In fact, to accomplish the task of finding galaxies through instance segmentation, there are various ways to construct a dataset. One of the most direct methods is to label galaxies in the actual telescope observational data and use the label masks as ground truth. This approach has two execution strategies: one is to use manual labels (similar to the process in many other instance segmentation tasks). Although simple and direct, this requires a certain amount of human labor for labeling and is not easily scalable. And manual labeling is always required each time when applying this machine-learning method to a new telescope. The second strategy is to use existing methods (e.g. template function method used by ALFALFA) for labeling, but this will cause a certain degree of `distortion' in the ground truth. This is because the recall and accuracy of existing methods are not $100\%$, which can lead to machine learning models becoming `similar to existing methods' after training.
    
        Besides using real observational data, another way to construct a dataset is to simulate data, such as using simulated galaxy data with real RFI, using real galaxy signals with simulated RFI, or using both simulated galaxy data and simulated RFI with noise background, etc.
        
        In our work, we used simulated galaxy data and simulated RFI. The reason is, first, the differences between our simulated galaxy signals and the real galaxy signals are minimal, so it is feasible to use simulated galaxy data as a substitute for real galaxy data. Second, it is difficult to search for and label galaxies directly in the TOD data, which is also the reason why we used simulated RFI. It is challenging to separate `pure' RFI without astronomical source signals from real data.
        Since the galaxy and RFI data are separately simulated, we can accurately and conveniently label them, which greatly assists us in our subsequent work.
        
        It is worth mentioning that the construction of the dataset is very flexible. For example, if we want the model to have the ability to identify galaxies among specific RFIs, we can add these particular RFIs (either simulated or real signals) to the original data, allowing the model to `learn' the ability to identify this type of RFI as interference. Furthermore, by adjusting the proportion of faint sources and bright sources in the dataset, the model can be more inclined to identify faint or bright source signals. The dataset construction method depends on the researcher, but all operations should be performed while ensuring the data is as realistic as possible.

    \subsection{Model Generalization and Potential Improvements}
    
        Although our simulation can obtain observational data for all 19 beams of the FAST telescope for HI galaxies, we only used data from one beam for the final training. This is because, in practice, though the response of different beams to the same signal is actually one of the bases for distinguishing galaxies and RFIs, we have not found a convenient way to simulate the same RFI received by different beams. Using the same RFI data for all beams may cause some errors.
    
        To further train the model, multi-beam data (e.g., FAST's 19-beam) can be used, inputting different beam data as different channels of two-dimensional data into the network. This allows the model to learn the response information of all beams for the same source and better search for galaxy signals.
    
        In addition, in the real observational data, besides RFI, other influences such as standing waves and bandpass of the system will have impacts on the search results, which means that the real data may be more complex than our simulations and require better detection capability of the model.
    
        Owing to the model's excellent generalization ability, one can also attempt to apply our network to the detection and extraction of signals from other astronomical sources. In fact, the PointRend Mask R-CNN can effectively perform instance segmentation tasks for numerous categories, while in our work it has only been used for a two-class (galaxies and interference) instance segmentation task. So our network can also be applied to data generated in other stages of telescope data processing for object detection and signal extraction (e.g., \citealt{riggi2023astronomical}).
    
        Additionally, when training the model, the weights of the different components in the total loss function can be adjusted to give the model a stronger `inclination'. For instance, increasing the weights of the $L_{cls}$ and $L_{box}$ components in the total loss function can make the model more inclined towards accurate recognition rather than precise segmentation. For a further saying, new network structures can be explored and incorporated into our model to enhance its capabilities for other missions.
    
        It should be noted that, limited by the accuracy of the numerical simulation, we currently consider galaxies with relatively high HI fluxes. Whether our models can perform better for galaxies with lower masses (galaxies fainter than those in our dataset) needs to be further investigated. Finally, we will apply our model to real observational data in our subsequent work, trying to perform galaxy searches in real data and comparing the results with other traditional methods.

\section{Conclusion}
\label{sec:conc}

    In our work, we constructed a Mask R-CNN network integrated with the PointRend method, aiming to find and extract galaxy signals in radio telescope observational data contaminated by RFI. We simulated the galaxy signals observed by FAST and the potential RFI impact as realistically as possible, and built a dataset based on this simulation for training and testing our network. We compared five different network architectures and chose the best-performing one, ultimately achieving precision and recall of $98.64\%$ and $93.59\%$, respectively. This demonstrates that our network can successfully accomplish the instance segmentation task of HI galaxy signals in TOD data.

    Moreover, thanks to the high-precision detailed performance of the PointRend method, our network can achieve more accurate segmentation when dealing with complex and subtle galaxy structures in astronomical images. We discussed the construction methods of the dataset and the possible generalizations and improvements of the model, believing that our network has excellent extensibility and can be applied to other scenarios.

    For the extraction of HI galaxy signals, although existing search algorithms have achieved some success in previous projects, there are still some drawbacks and challenges in practical applications. Our bold attempt to find galaxy signals using a deep neural network is an innovative application of machine learning methods to this task, which helps provide more reliable basic data for subsequent astronomical analyses and lays a better foundation for the next step of scientific research.

\begin{acknowledgements}
    We thank Long Xu and Dong Zhao for their useful discussions. 
    We acknowledge the support by National SKA Program of China, No.2022SKA0110100, the CAS Interdisciplinary Innovation Team (JCTD-2019-05). We also acknowledge the science research grants from the China Manned Space Project with NO.CMS-CSST-2021-B01.
\end{acknowledgements}

\bibliographystyle{raa}
\bibliography{bibtex} 

\end{document}